\documentclass{aastex631}
\usepackage{amsmath}

\shorttitle{Self-similar coronal outflows and reconnection}
\shortauthors{Uritsky et al.}

\newcommand{\kms}{km\,s$^{-1}$}

\newcommand{\corr}[1]{{{#1}}}

\begin{document}

\title{Self-Similar Outflows at the Source of the Fast Solar Wind: \\
A Smoking Gun of Multiscale Impulsive Reconnection?}

\correspondingauthor{Vadim Uritsky}
\email{uritsky@cua.edu}

\author{Vadim M.\ Uritsky}
\affiliation{Catholic University of America, 620 Michigan Avenue NE, Washington, DC 20064, USA}
\affiliation{Heliophysics Science Division, NASA Goddard Space Flight Center, Greenbelt, MD, 20771, USA}

\author{Judith T.\ Karpen}
\affiliation{Heliophysics Science Division, NASA Goddard Space Flight Center, Greenbelt, MD, 20771, USA}

\author{Nour E.\ Raouafi}
\affiliation{The John Hopkins University Applied Physics Laboratory, Laurel, MD 20723, USA}

\author{Pankaj Kumar}
\affiliation{Department of Physics, American University, Washington, DC 20016, USA}
\affiliation{Heliophysics Science Division, NASA Goddard Space Flight Center, Greenbelt, MD, 20771, USA}

\author{C.\ Richard DeVore}
\affiliation{Heliophysics Science Division, NASA Goddard Space Flight Center, Greenbelt, MD, 20771, USA}

\author{Craig E.\ Deforest}
\affiliation{Southwest Research Institute, 1050 Walnut Street, Suite 300, Boulder, CO 80302, USA}

\begin{abstract}
 
We present results of a quantitative analysis of structured plasma outflows above a polar coronal hole observed by the Atmospheric Imaging Assembly onboard the Solar Dynamics Observatory spacecraft. In a 6-hour interval of continuous high-cadence SDO/AIA images, we identified more than 2300 episodes of small-scale plasma flows in the polar corona. The \corr{mean} upward flow speed \corr{measured by the surfing transform technique \citep{uritsky2013}} is estimated to be 122 $\pm$ 34 \kms, which is comparable to the local sound speed. The typical recurrence period of the flow episodes is 10 to 30 minutes, and the mean duration and transverse size of each episode are about 3--5 min and 3--4 Mm, respectively. The largest identifiable episodes last for tens of minutes and reach widths up to $40$ Mm. For the first time, we demonstrate that the polar coronal-hole outflows obey a family of power-law probability distributions characteristic of  impulsive interchange magnetic reconnection. 
\corr{Turbulent} photospheric driving  may play a crucial role in releasing magnetically confined plasma onto open field. The estimated occurrence rate of the detected 
\corr{self-similar}
coronal outflows is sufficient for them to make a dominant contribution to the fast-wind mass and energy fluxes and to account for the wind's small-scale structure.
  
\end{abstract}


\section{Introduction} \label{sec:introduction}

Small-scale plasma outflows in magnetically open regions of the solar corona have been observed for a long time. In many open geometries, including mid-latitude plumes \citep{raouafi2014} and plumelets \citep{uritsky2021}, polar plumes \citep{raouafi2008}, erupting spicules \citep{sterling2016}, and reconnection-driven jets and jetlets \citep{raouafi2023, kumar2022, kumar2023, sterling2023}, the plasma outflow tends to organize itself into fine-scale filamentary structures constrained by the expanding magnetic field. The characteristic transverse sizes of these structures have been shown to lie in the range 2--10 Mm, with typical lifetimes on the order of 5--10 minutes \citep{uritsky2021, kumar2022}. The upward speed of the structured outflow is comparable to the sound speed, so waves and bulk motions could coexist -- see, e.g., \citet{uritsky2013} and references therein. Numerical simulations indicate that the structured outflow can be driven by low-lying magnetic reconnection releasing and accelerating the frozen-in plasma \citep{pariat2009, karpen2017, wyper2018, drake2023}. The resulting jet is composed of a relatively dense region near the coronal base exhibiting significant pressure and density modulations, an intermediate region at a higher altitude filled with Alfv\'enic and compressional perturbations, and a purely Alfv\'enic front well ahead of the slower, denser mass flow \citep{uritsky2017}. Signatures of interchange reconnection and the resultant compressible outflow in the lower corona have been confirmed in coronal-hole observations \citep{kumar2022, mason2022}. 

A key open question is the role of the small-scale coronal outflows in the formation of the large-scale solar wind. It has been argued \citep{raouafi2023} that the former could be building blocks of the latter by providing enough plasma mass, momentum, and kinetic energy to fuel the transient component of the solar wind. 
\corr{It has also been suggested \citep{wang2022} that quasi-continuous interchange reconnection could drive the ambient solar wind by a combination of nanoflare-like plasma heating \citep{klimchuk2006,klimchuk2015} and MHD-wave deposition of momentum and kinetic energy \citep{cranmer2019,zank2021}.} 
Other estimates suggest that the occurrence frequency of the small-scale jets is at least high enough to cause significant perturbations in the outflowing plasma \citep{kumar2022, drake2023}.  In either scenario, the reconnection-driven flows definitely generate upward propagating discontinuities \citep{uritsky2017, roberts2018}, which might give rise to the sudden rotations of the interplanetary magnetic field known as switchbacks \citep{horbury2020, kumar2022, kumar2023, bale2021, bale2023}. The switchbacks have been observed in abundance by Parker Solar Probe \citep{fox2016,raouafi2023b}, revealing the highly structured nature of the solar wind.

In this paper, we report results of a quantitative statistical survey aimed at evaluating the contribution of the structured transient plasma outflows to the bulk solar wind. We use high-resolution extreme ultraviolet (EUV) images of the dynamic solar atmosphere above a polar coronal hole to identify more than 2300 individual small-scale jets and measure their spatial scale, duration, upward speed, and entrained mass. Our analysis demonstrates that the structured plasma outflows obey a family of multiscale probability distributions characteristic of turbulent impulsive reconnection, which may play a crucial role in releasing the heated low-coronal plasma into the solar wind.

\section{Data and methods} \label{sec:methods}

We have analyzed a sequence of 1,801 EUV images collected by the Atmospheric Imaging Assembly onboard the Solar Dynamics Observatory \citep{lemen2012,pesnel2012}. The images covered a 6-hour interval between 2021-04-28 00:00:09 and 2021-04-28 05:59:57, with the cadence of 12 s, spatial resolution 1.5\arcsec, and plate scale 0.6\arcsec, or 435 km. We used 171 {\AA} images dominated by Fe IX emission at a characteristic temperature of $6.3 \times 10^5$~K observed in the quiet corona and the upper transition region. The selected time interval is sufficiently long to obtain statistically representative signatures of the studied flow structures, which are omnipresent in the open-field corona. 

\begin{figure}
\begin{center}
\includegraphics[width=8.5 cm]{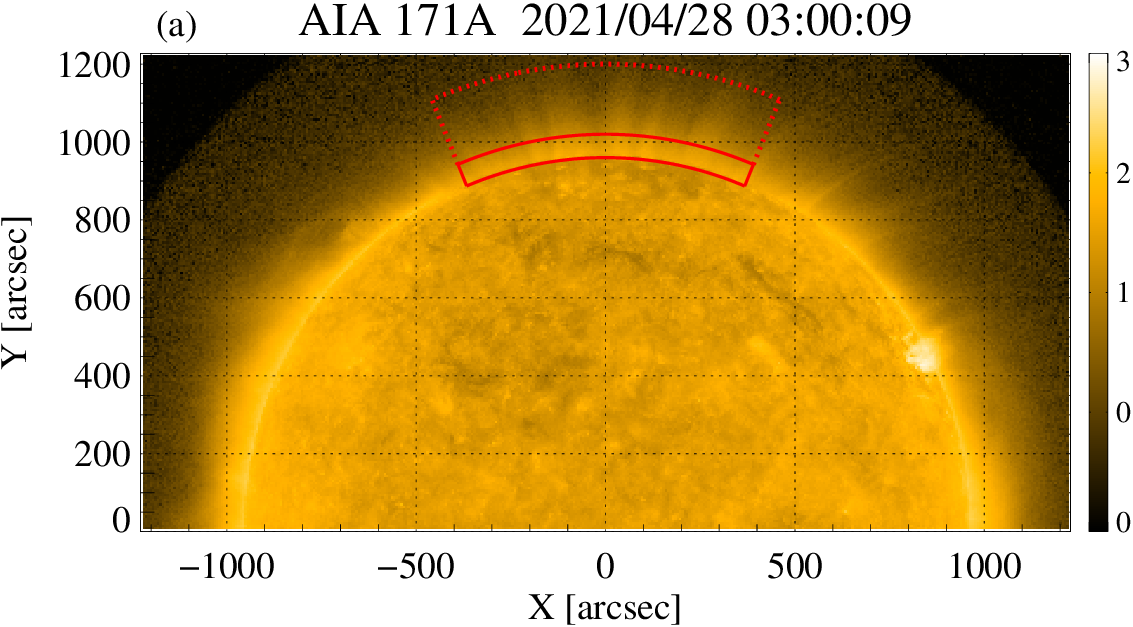}

\includegraphics[width=8.5 cm]{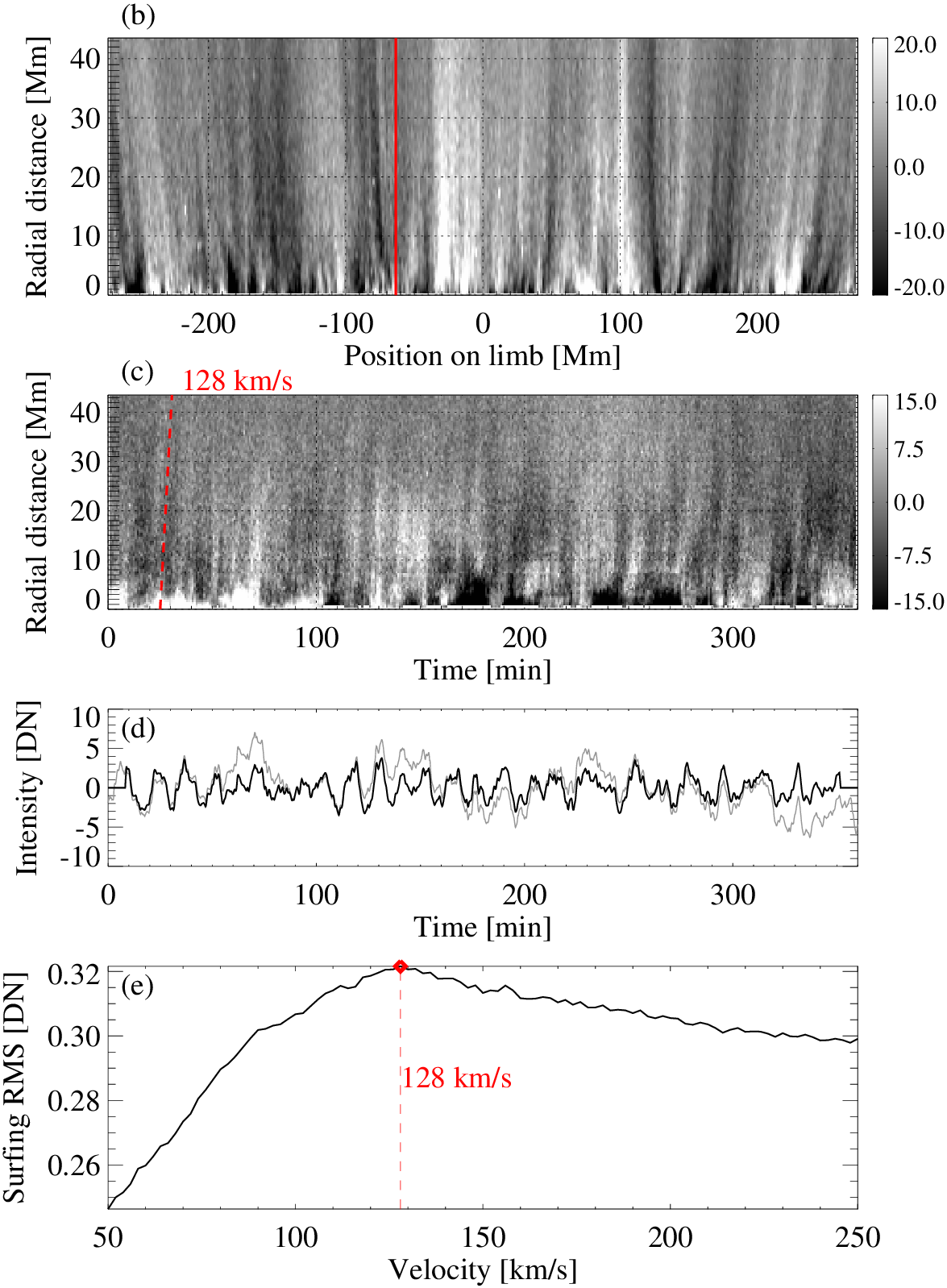}
\caption{\label{fig_flows} (a) An SDO/AIA 171 \AA\ image  in the middle of the studied 6-hour time interval, with the region of interest bounded by the red lines. The solid line shows the main ROI used for the statistical survey; the dotted line is an ROI with an extended upper boundary used for the speed measurements based on the ST method. A base-ten logarithmic brightness scale is used to enhance the low-intensity structure above the limb. (b) Plane-spherical coordinate representation of the main ROI on panel (a) after radial trend removal. The highly filamented texture of the outflowing plasma is easily seen. (c) \corr{One of 1257 studied time-distance plots} representing time evolution of the lower corona along the radial slit marked with a \corr{red} dashed line on panel (b).  \corr{The gray scales on panels (b) and (c) are calibrated in linear AIA DN units.} A spatiotemporal direction corresponding to the propagation velocity measured by the surfing analysis is shown with a dashed line. (d) The maximized surfing signal $S(t,u=v)$ fulfilling the velocity resonance condition, before and after temporal detrending (gray and black lines, respectively). (e) The RMS value $S_{RMS}$ of the surfing signal $S(t,u)$ for the time-distance plot in panel (c), showing a well-defined peak at $u=v$.}
\end{center}
\end{figure}

Our investigation was focused on the quasi-steady plasma outflow observed above the north polar limb and launched from the polar coronal hole. The region of interest (ROI)  covered a $45^\circ$ limb sector centered at the north pole (Figure \ref{fig_flows}(a)). Two ranges of altitude above the limb (4.35 Mm--174 Mm and 4.35 Mm--43.5 Mm) were used to study, respectively, the velocity and the occurrence statistics of the outflows as explained below. The images were transformed into plane-polar coordinates to simplify the analysis of quasi-radial structures, with a position-angle bin size of $6.25 \times 10^{-4}$ radians (0.0358$^\circ$) matching the native AIA resolution at the base of the ROI and exceeding it by a factor of 1.26 at the upper ROI boundary, and a radial position bin equal to the original AIA resolution. After transforming the images into polar coordinates, we subtracted the time-averaged local radial decay of the image intensity at each azimuthal position. The resulting detrended polar-coordinate arrays were smoothed using a $3\times 3$-pixel boxcar window to reduce  binning errors and pixel noise (Figure \ref{fig_flows}(b)), and transformed into a set of 1257 time-distance arrays representing plasma propagation dynamics at each location along the limb (Figure \ref{fig_flows}(c)). 

Propagating disturbances in each time-distance array were identified using the surfing transform (ST) technique \citep{uritsky2013}. This technique is designed to identify wave signals in time–distance plots with low signal-to-noise ratios, \corr{and it has been used successfully to measure parameters of remotely observed flows and waves \citep{uritsky2009, keiling2012, uritsky2013, uritsky2021, kumar2022, raouafi2023, mondal2023}.} The method is based on an analysis of a velocity-dependent surfing signal $S(t, u)$ obtained by averaging the time-distance array along the spatiotemporal defined $u$, starting form time $t$. The dynamic range of the signal is maximized when the assumed surfing velocity $u$ matches the true propagation velocity $v$, in which case the averaging is performed in the reference frame of the propagating disturbance. \corr{The main factor limiting the accuracy of the ST method is the the nonstationarity of the signal. For quasi-periodic flows such those studied here, the ST uncertainty is typically in the range 2-5 km s$^{-1}$ \citep{uritsky2013, mondal2023}, which is an order of magnitude smaller than the spatial variability of the flow velocity reported in \S \ref{sec:results}.} Figure \ref{fig_flows}(d) shows the maximized surfing signal meeting this condition for the time-distance plot in Figure \ref{fig_flows}(c) at $v=128$ \kms. The quasi-periodic nature of the disturbance and its nonlinear waveform are evident. Figure \ref{fig_flows}(e) shows the time-averaged root-mean-square value $S_{RMS}$ of the surfing signal measured plotted as a function of $u$, with a clear-cut peak at $u=v$. The peak broadens in the presence of long-term trends, which have to be removed for more accurate results.

The ST analysis illustrated by Figure \ref{fig_flows} was applied to all positions along the limb in the studied data set, which allowed us to estimate the spatially dependent propagation speed characterizing the plasma outflow. Next, the spatiotemporal coherence between the propagation dynamics at neighboring locations was investigated to identify the individual jet events and evaluate their amplitude and scales. Finally, we looked at the statistics of the obtained outflow parameters in the context of the large-scale solar wind and the reconnection-driven jet mechanics. 

\section{Results and discussion} \label{sec:results}

\begin{figure}
\begin{center}
\includegraphics[width=8.5 cm]{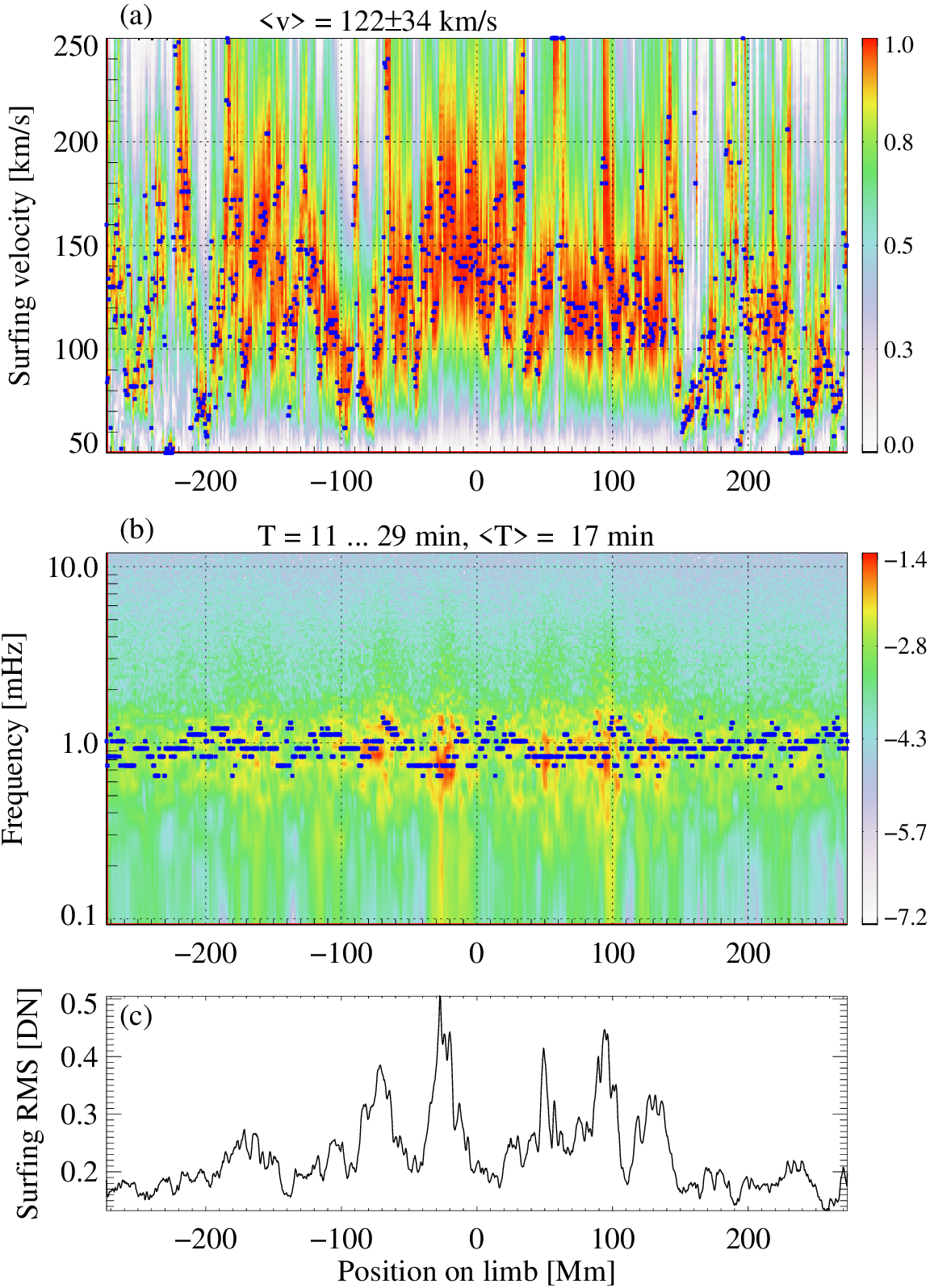}
\caption{\label{fig_surf} ST analysis of the coronal outflow speed. (a) Speed dependence of the surfing RMS for each location on the limb. Each vertical slice represents a $S_{RMS}(u)$ dependence such as that plotted on Figure \ref{fig_flows}(e), with the blue dot showing the position of the resonance maximum used to identify the propagation speed. (b) Fourier power spectra of the resonance surfing signals for different limb positions. The blue dots show the spectral peaks used to evaluate the dominant frequency. Temporal trends longer than 20 min were removed for a more accurate velocity estimation. (c) Resonance value $S_{RMS}(u=v)$ representing the characteristic amplitude of the propagating quasi-periodic disturbance at each location.}
\end{center}
\end{figure}

Figure \ref{fig_surf} shows the results of the velocity analysis of the coronal outflow across the studied region above the limb. The top panel visualizes the velocity dependence $S_{RMS}(u)$ of the surfing RMS for each of the 1257 studied limb positions. The blue dots mark the location of the RMS peak on each curve used to evaluate the resonance surfing velocity $v$. The peak values found at or near the edges of the studied $u$ interval (50 or 250 \kms) were excluded from further analysis as unreliable. \corr{ The one-sigma range of the flow speed variability is 88 to 156 \kms, } with a mean value of 122 {\kms} that is roughly the local sound speed. The observed range of speeds is somewhat lower than the characteristic speed of jetlet activity reported previously for the same coronal region \citep{raouafi2023} and is systematically below the typical speed ($\approx$ 200 \kms) in mid-latitude coronal plumelets \citep{uritsky2021}. That the detected outflows are quasi-periodic is suggested by the power spectral analysis of the resonant surfing signals (Figure \ref{fig_surf}(b)). The period of the propagating structures varies across limb positions in the range of 10 to 30 min, with an average of 17 min, which is roughly consistent with recent observations of propagating disturbances in plumes within polar coronal holes \citep{cho2021}. Due to the applied temporal detrending, it is possible that the image sequence contains time scales that are longer than those appearing on the spectrogram. However, it is evident that no higher frequencies are present in the studied polar coronal region. In particular, there are no signs of the ubiquitous 3--5 min $p$-mode oscillations accompanying the small-scale outflows in the plumelets and jetlets. This is likely because the plume studied here is faint and rooted in weak network/plage flux concentrations, whereas the plumes studied in our previous works \citep{uritsky2021, kumar2022, kumar2023} are bright and rooted in strong network/plage regions. Jet activity at the base of plumes is a function of the strength of the associated flux concentrations, a dependence that requires further investigation. 

The degree of modulation of the optical intensity by the propagating disturbances is quite low, as evidenced by Figure \ref{fig_surf}(c) showing peak $S_{RMS}$ values at each position. The flow intensity varies between 0.1 and 0.5 AIA DN units, which is a small fraction of the background intensity of the same coronal region (see Figure \ref{fig_flows}(b)).

\begin{figure}[ht]
\begin{center}
\includegraphics[width=8.5 cm]{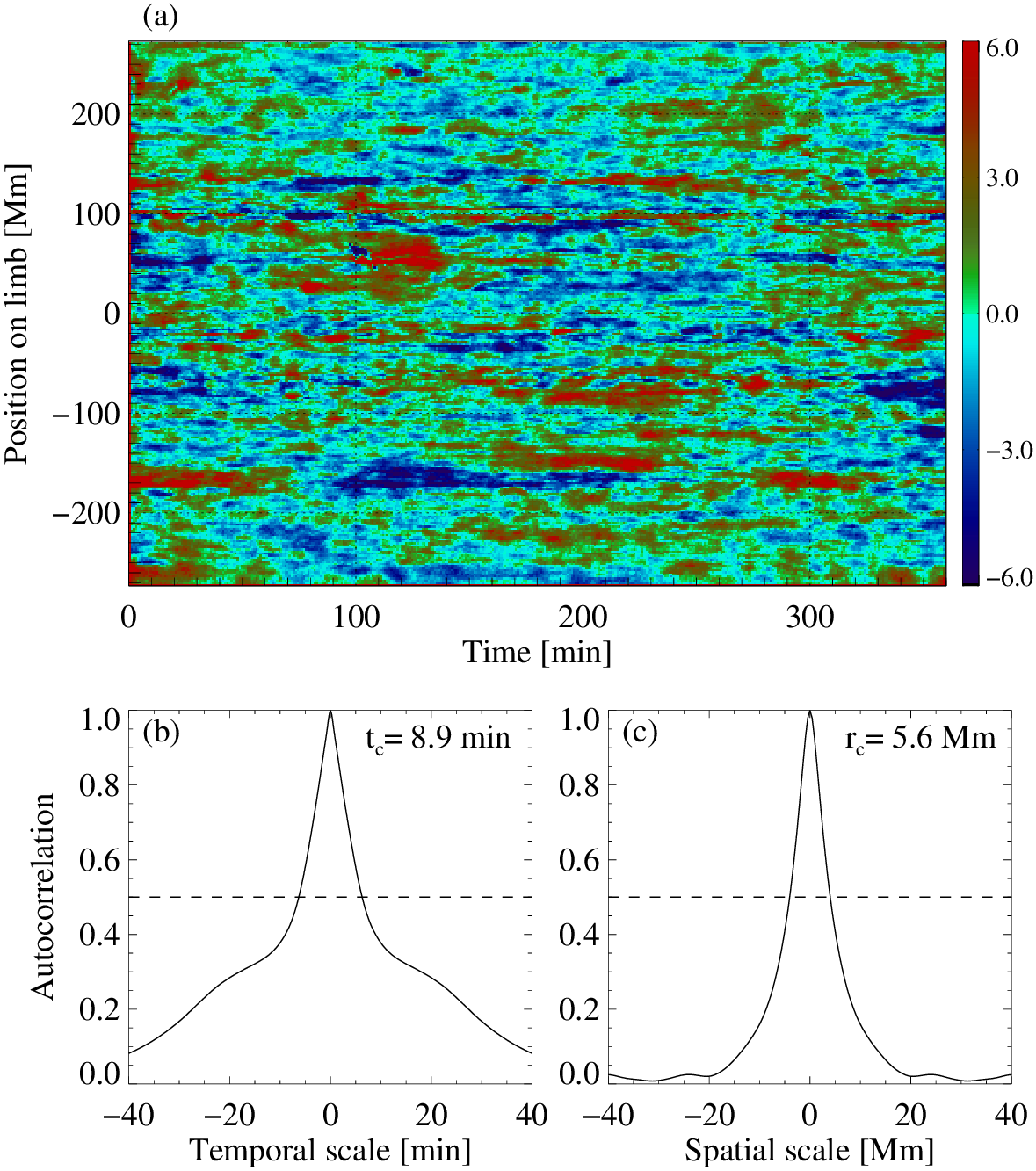}
\caption{\label{fig_corr} Top: Spatiotemporal map of the transient outflow in the studied region constructed using the ST technique. The compact blobs shown in red represent starting times and locations of the quasi-periodic propagating fronts (see text for details). Bottom: Temporal and spatial autocorrelation functions of the surfing signals; the provided characteristic scales represent the 50$\%$ levels and assume an exponential decay.}
\end{center}
\end{figure}

Besides being a sensitive tool for velocity analysis, the ST technique provides a natural means to detect small-scale outflows appearing at different times and limb locations. If a single small-scale jet spans more than one azimuthal position, the surfing signals at all positions within the jet should be time-correlated, reflecting coherent plasma propagation. By isolating clusters of such correlated pixels and measuring their spatial and temporal sizes, one can therefore evaluate the spatial and temporal scales of the small-scale jets. Figure \ref{fig_corr} shows a spatiotemporal map $S_{RMS}(t, x, u=\langle v \rangle)$ of the flow activity in the studied coronal region, where $t$ is the time, $x$ is the position along the limb, and $\langle v \rangle = 122$ \kms~is the average outflow velocity reported in Figure \ref{fig_surf}(a). Each horizontal row of pixes is a surfing signal at a given limb position. No temporal high-pass filtering has been applied to this plot, to preserve all relevant time scales. The clustered groups of red pixels are the coherent crests of the surfing signals; likewise, the  clusters of blue pixels are the coherent troughs. Each blob can be interpreted as an individual propagating front described by a certain transverse size and duration. 

Based on Figure \ref{fig_corr}(a), the outflows clearly involve a wide range of spatial and temporal scales, making it difficult to pinpoint a single combination of scales that could be used as a hallmark of a specific morphology, such as a jetlet or a plumelet. Some flow bursts are only 2--3 azimuthal pixels ($\sim 1$ Mm) wide, while other encompass 20--30 pixels ($\sim$ 10--15 Mm). Some flow bursts last for several minutes only, while others are sustained for up to 30--50 minutes. While this multiscale behavior cannot be described by precisely defined scales, it can be characterized statistically by a set of probability distributions. To perform such a statistical analysis, we decomposed the complex flow activity pattern shown in Figure \ref{fig_corr}(a) into a set of discrete flow events by performing a spatiotemporal segmentation. To make this segmentation fully objective, we based our algorithm on the intrinsic correlation scales of the pattern and an automatically optimized detection threshold, as described below.

\begin{figure}
\begin{center}
\includegraphics[width=8.5 cm]{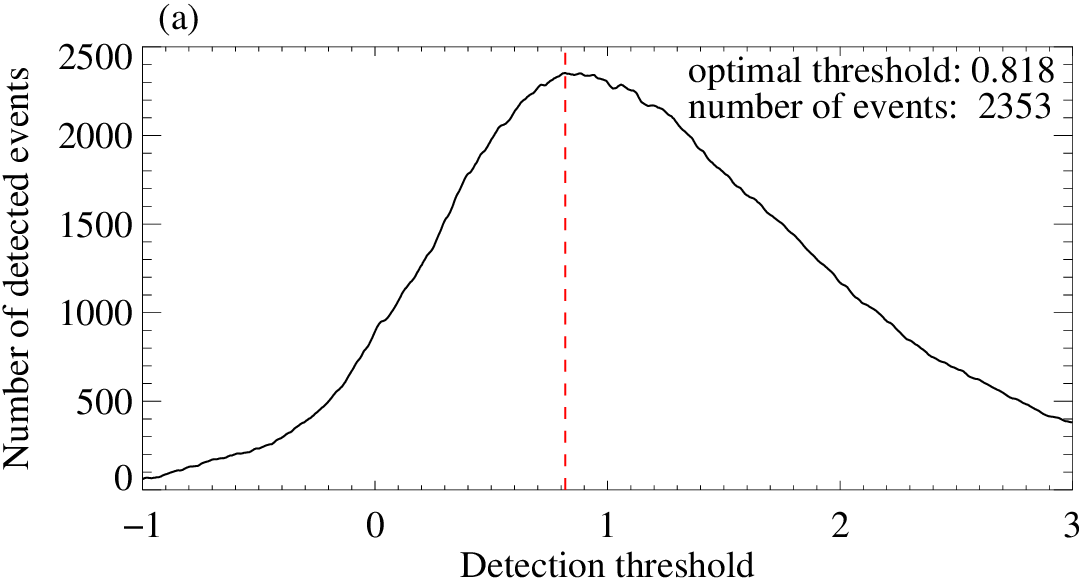}

\includegraphics[width=8.5 cm]{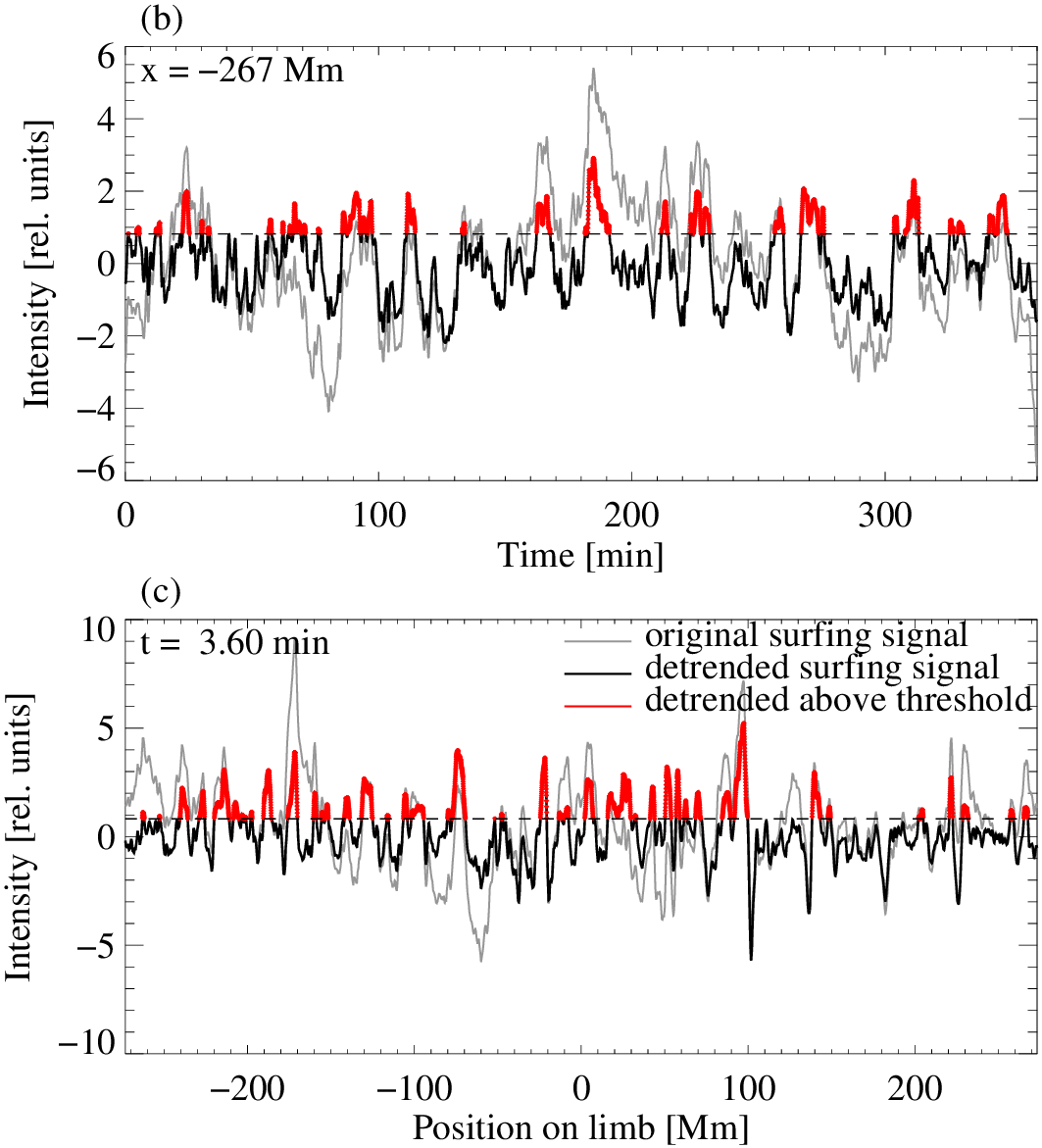}

\includegraphics[width=8.5 cm]{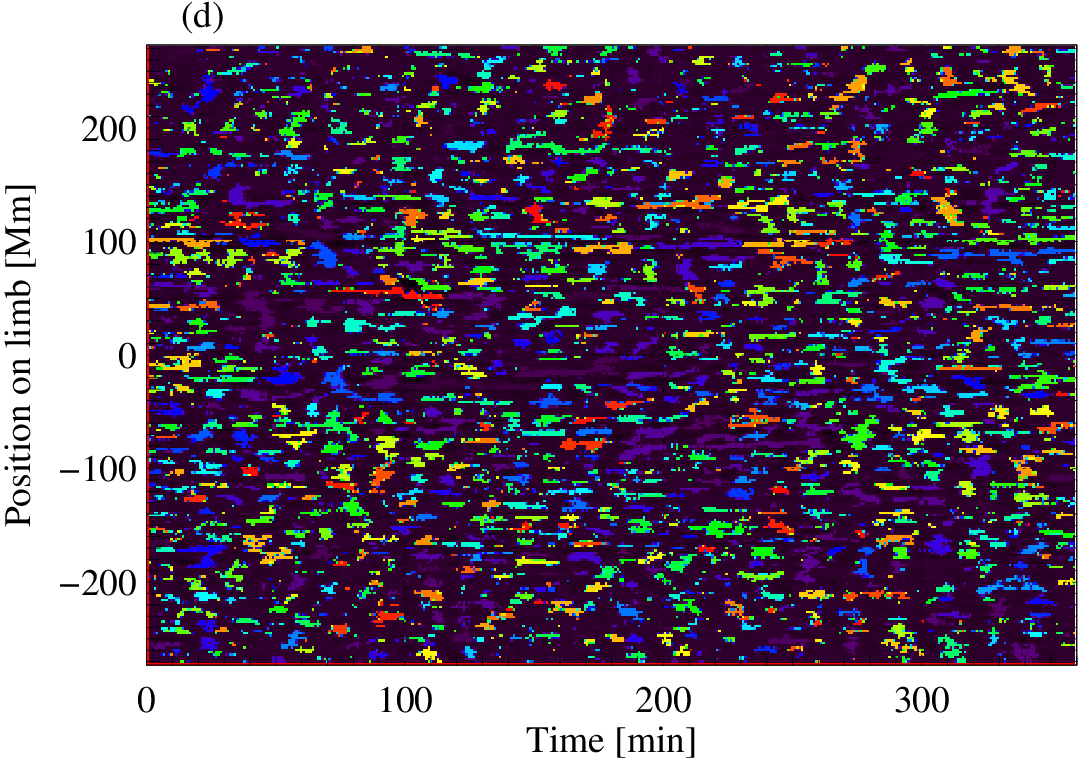}

\caption{\label{fig_thresh} (a) The number $N_f$ of distinct identifiable flow fronts on the detrended surfing map versus the detection threshold. The optimized threshold (0.818) resulting in the largest number of identified flow events (2353) was used for the subsequent spatiotemporal event detection. (b,c) Example of temporal and spatial detrending of the surfing flow map $S_{RMS}(t, x)$ based on the estimated correlation scales. (d) The processed flow map showing the detected flow events marked with random color. Only the events involving more than 10 spatiotemporal pixels were plotted and used for the statistical survey presented in Figure \ref{fig_stat}, \corr{ since smaller events are not informative due to the applied image smoothing.}}
\end{center}
\end{figure}

\begin{figure}
\begin{center}
\includegraphics[width=11.0 cm]{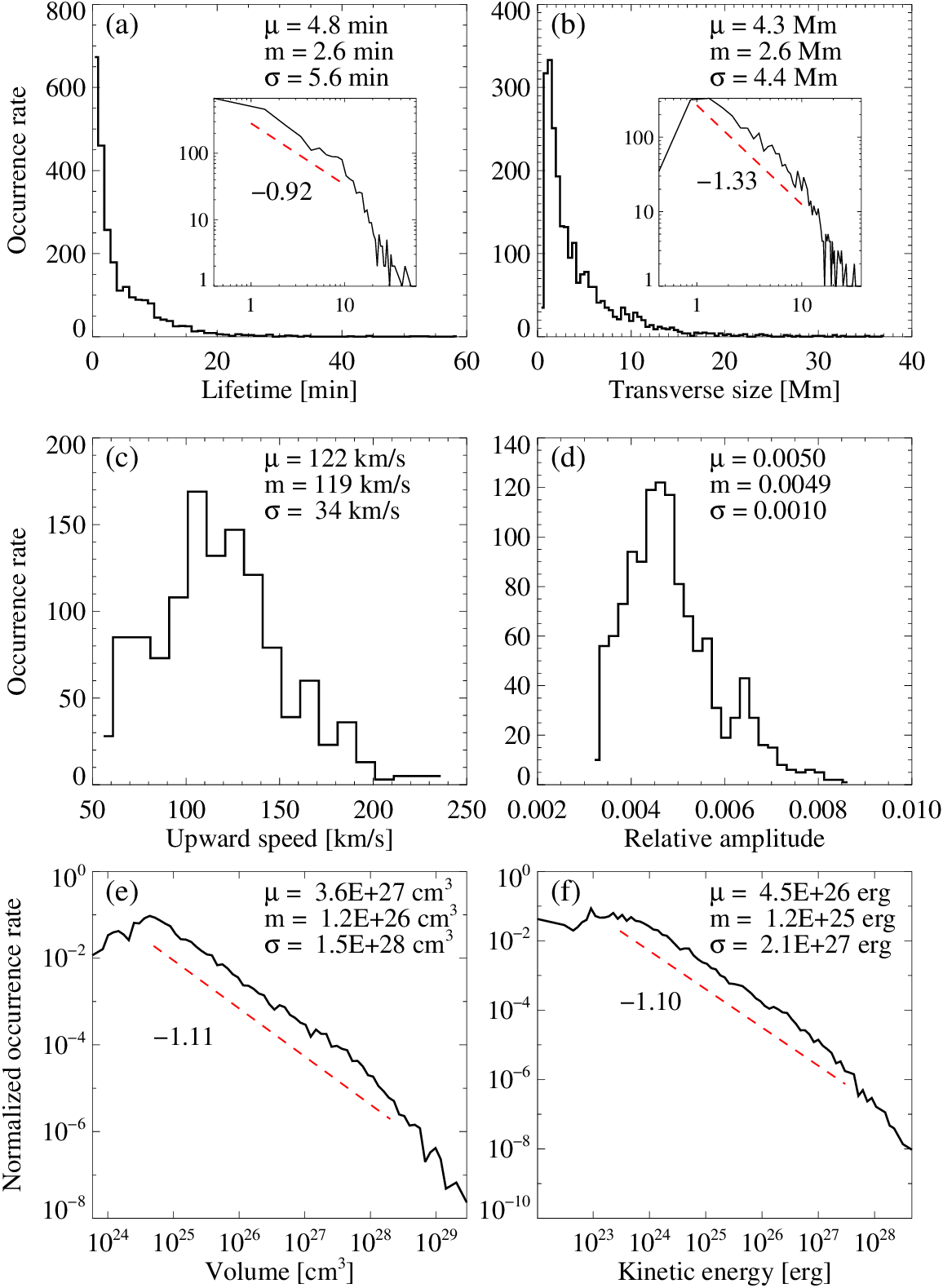}
\caption{\label{fig_stat} Probability histograms of the detected transient flow events over the event: (a) lifetime $T$; (b) transverse flow size $L$; (c) upward flow velocity $v$; (d) relative emission intensity $\delta I / I_0$; (e) transported flow volume $V$; and (f) kinetic energy $K$ of the flow. \corr{Mean ($\mu$), median ($m$), and standard deviation ($\sigma$) values are provided for each parameter}. The dashed red lines show the log-log slopes of the $T$, $L$, $V$, and $K$ distributions, supplemented by the evaluated power-law exponent values.}
\end{center}
\end{figure}

Panels (b) and (c) of Figure \ref{fig_corr} show temporal and spatial autocorrelation functions of the $S_{RMS}(t,x)$ array. We evaluated the correlation scales ($t_c$  and $r_c$, respectively) at the half-maximum level; the values are provided on the plots. Next, we removed the trends from the $S_{RMS}(t,x)$ array by subtracting its boxcar-averaged version, with the sliding window size in the $t$ and $x$ directions set to $2 \, t_c$ and $2 \, r_c$, respectively. The resulting detrended map contains compact flow events that can be isolated by applying a constant detection threshold. The number $N_s$ of the detected events depends upon the threshold. To find the threshold that produces the largest population of flows, we varied the threshold in the range from $-$1 to 3 and located the position of the peak with the maximum $N_f=2353$ (Figure \ref{fig_thresh}(a)). The detection threshold of 0.818 AIA DN units corresponding to this peak was used for the subsequent analysis. \corr{Additional tests have confirmed that the power-law components of the flows distributions reported  on Figure \ref{fig_stat} are insensitive to the particular choice of the threshold, although the range of the scaling behavior gradually contracts if the threshold exceeds the optimal value.}

Figure \ref{fig_thresh}(b) shows an example of a surfing signal computed at the limb position $x=-267$ Mm before (gray line) and after (black line) the detrending procedure. The signal values above the optimized detection threshold 0.818 are marked with red color. Figure \ref{fig_thresh}(c) presents a similar plot representing an azimuthal profile of the flow map at $t=3.6$ min measured from the beginning of the studied time interval. 

Figure \ref{fig_thresh}(d) shows a map of the detected flow bursts using the methodology described above. The individual flow events detected automatically as contiguous pixel clusters are marked with  random colors; the black background is below the detection threshold and represents the ambient plasma. Two aspects of the analysis should be noted before moving forward with the statistical analysis of the constructed flow population. First, the detected events lack the largest and longest-duration scales exhibited by the flows in the non-detrended, \corr{non-thresholded} flow map (Figure \ref{fig_corr}(a)). This bias is an unavoidable price paid for segmenting a continuous outflow pattern. \corr{It implies that both the upper scaling cutoffs and the net contribution of the flow material and energy to the solar wind could be larger than those revealed by the flow statistics.}
Second, we had no unambiguous way to verify the statistical homogeneity of the obtained set of events, although its resulting statistics presented below speak in favor of a reasonably uniform and representative sample. 

The probability distributions of the flows over the lifetime $T$ and the transverse size $d$ still appear remarkably multi-scale, spanning over nearly two orders of magnitude (Figure \ref{fig_stat}(a,b)). The mean ($\mu$) and median ($m$) values indicated on the histogram panels are biased by the smallest events, which have the highest occurrence rate. The reported values are approximately consistent with earlier published estimates, but the fact that the power-law form of the two distributions continues down to the smallest scales resolved by our analysis could be an indication that even smaller and shorter flow bursts are present in the corona. \corr{The standard deviation $\sigma$ of both $T$ and $d$ parameters is of the order of the mean value, reflecting the wide statistical spread associated with heavy-tailed distributions.} The log-log slopes (dashed red lines in the insets of panels (a) and (b)) of the power-law portions of the distributions unaffected by detrending are greater than $-2$, which suggests that the mean flow scales could be weighted toward the largest, rather than the smallest, outflow events. \corr{The small-scale size distribution cutoff ($\approx 1.5$ Mm) is caused by the $3\times3$-pixel image smoothing. It does not reflect the smallest physical scale of the flows, which could be lower.}

Compared to the spatiotemporal scales, the velocity distribution of the flows (Figure \ref{fig_stat}(c)) is relatively compact, with the mode and the median roughly consistent with the mean as expected for Gaussian statistics. A possible secondary peak around 70--80 \kms~could be a signature of a separate population of flows, pending its statistical significance being confirmed in future studies involving larger sets of events.

Figure \ref{fig_stat}(d) presents the probability distribution of relative flow intensities obtained by dividing the peak surfing intensity $\max(S_{RMS})$ measured at each position on the limb by the time- and height-averaged AIA brightness computed for the same azimuthal position. This ratio varies between about $3\times 10^{-3}$ and $8\times 10^{-3}$, with a mean of $5\times 10^{-3}$. The interpretation of this weak EUV emission signature in terms of plasma density modulation $\delta n$ caused by the flow relative to the unperturbed background density $n_0$ requires a line-of-sight (LOS) integration of the luminous plasma material in the compact flow versus the ambient corona. Assuming similar filling factors in the two plasma systems, the ratio of the emission signals can be expressed as $\delta I / I_0 \sim (L/D) (\delta n^2/ n_0^2)$, where $L$ and $D$ are respectively the characteristic transverse size of the flow and the characteristic LOS depth of the corona at a given distance from the limb. The assumption of similar filling factors is an important uncertainty \citep{deforest1991}, because the observed jets could, in principle, have a smaller filling factor than the surrounding corona. \corr{This follows, e.g., from the comparison of plume basal electron densities based on density sensitive spectral line ratios \citep{warren1999} versus the EUV photometry \citep{walker1993}.} We defer verification of this possibility to future work. Using the available empirical radial density profile models \citep{cranmer1999, guhathakurta1999}, we evaluated the e-folding $D$ value to be on the order of 0.7 solar radii ($\approx 500$ Mm) for the range of altitudes covered by the studied solar region. Based on our flow analysis,  $\langle \delta I/I_0 \rangle \approx 5\times 10^{-3}$ and $\langle d \rangle \approx 4$ Mm, so the relative density perturbation caused by the typical flow $\delta n / n_0 = \sqrt{ (D/L) (\delta I/I_0)} \approx 0.6$ and the total density of the flow $n \approx 1.6\, n_0$. Therefore, despite its minuscule emission footprint, the detected flows can cause a substantial local density enhancement. \corr{Importantly, a lower filling factor for the observed flows would result in an even more significant $\delta n/n_0$.}

 The volume of plasma transported by each transient flow event can be evaluated as $V = v\,d^{\,2}\,T$. Since both $d$ and $T$ have wide probability distributions with distinct power-law ranges, the distribution of volumes (Figure \ref{fig_stat}(e)) exhibits an even broader interval of power-law behavior spanning over at least four orders of magnitude. The distribution of kinetic energy in the flows $K = (1.6\,n_0) \, m_p \, v^{\,2} \, V \,/ 2$, in which we used a typical background number density $n_0 \approx 5\times 10^8$ cm$^{-3}$ near the base of the corona \citep{raouafi2008} for pure hydrogen composition, also shows a broad interval of power-law decay (Figure \ref{fig_stat}(f)). As for the lifetime and size distributions, the power-law exponents describing the statistics of $V$ and $K$ are well above $-2$, implying that the characteristic values of the transported plasma volume and energy are weighted toward large events. \corr{The standard deviations of $V$ and $K$ parameters are several times greater than the corresponding mean values, as expected for broad-band power-law statistics with a shallow log-log slope.} 

 The obtained probability distributions of multiscale polar jets provide new evidence for the reconnection-driven mechanism of the coronal plasma outflow. The power-law volume distribution $p(V) \propto V^{-\tau_V}$ can be derived directly from the jet size distribution $p(L) \propto V^{-\tau_L}$ under the assumption that the outflow speed is independent of the width $L$ of the ejected flow observed at a distance from the reconnection site, which seems plausible. The power-law form of $p(T)$ implies that the lifetime $T \propto L^{\,D_T}$, and therefore $V(L) \propto  L^{\,2+D_T}$, where $D_T$ is the lifetime geometric exponent \citep{uritsky2013a}. From the probability conservation relation $p(V)dV = p(L)dL$ we obtain
\begin{equation}
\label{eq:scaling}
    \tau_V = \frac{\tau_L + D_T + 1}{D_T +2}.
\end{equation}
 The substitution of the measured values $\tau_L\approx 1.33$ and $D_T \approx 1$ into (\ref{eq:scaling}) results in $\tau_V \approx 1.1$, which is consistent with our statistical analysis. Since $K\propto V$, the same conclusion applies to the kinetic energy distribution exponent $\tau_V$, also in agreement with our observations. 

 The validity of the theoretical relation (\ref{eq:scaling}) indirectly confirms that the acceleration mechanism underlying the detected outflows is rooted in reconnection physics. \corr{The observed self-similarity of the flow pattern is likely imposed by the spatial complexity of the photospheric source regions \citep{uritsky2013a, uritsky2014}, which contain embedded bipoles spanning a wide range of scales \citep{karpen2017, kumar2021}.} In this context, it is worth noting that the slopes of the $p(V)$ and $p(K)$ distributions are roughly consistent with the power-law slopes describing the distribution of volumes and dissipated energies in fragmented pieces of a full-size coronal jet in a high-resolution 3D MHD simulation (see Figure 13 of \citet{uritsky2017}), which leaves the  possibility that at least some of the propagating disturbances detected here are produced by nonlinear instabilities \citep{yuan2019} in a large-scale turbulent jet rather than by stand-alone small-scale jetlets.
 
Whether the small-scale flows constituting the majority of the studied statistical population make a significant contribution to the solar wind depends upon their occurrence rate and the amount of plasma mass and kinetic energy injected into the corona. Using the mean volume and energy estimates shown in Figure \ref{fig_stat}(e) and (f), the net volume flow rate by all the flows in the studied polar region is $N_f \left\langle V\right\rangle / T_{obs} \approx 4 \times 10^{26} $ cm$^3$ s$^{-1}$, where $N_f = 2353$, $T_{obs} = 6 \times 3600$ seconds, and $\left\langle \, \right\rangle$ denotes ensemble averaging. The obtained volumetric speed translates into a particle flow rate of about $2 \times 10^{35}$ protons s$^{-1}$ (using the same nominal number density as in the $K$ calculation), and a mass flow rate $\approx 3 \times 10^{11}$ g s$^{-1}$. In turn, the total kinetic-energy rate (the kinetic power supplied by all detected flows) is $N_f \left\langle K \right\rangle / T_{obs} \approx 5 \times 10^{25} $ erg s$^{-1}$. 
\corr{The propagated standard errors of the provided estimates cause variations of 10\% or less from the mean values, although the true statistical uncertainties accounting for data nonstationarity, event detection errors, and observational constraints could be much larger.}

The photospheric surface area of the 45$^{\circ}$ spherical sector defining of the studied ROI (Figure \ref{fig_flows}(a)) is about $4\%$ of the surface of the Sun. If the cumulative footpoint area of the open-field coronal morphologies supporting structured plasma outflow were only 3 times larger, the particle flow rate associated with studied transient flows would be sufficient to generate the net solar-wind particle loss rate $6\times 10^{35}$ protons s$^{-1}$ predicted for the entire Sun \citep{wang2016a, wang2020}. The kinetic energy rate produced by the flows is consistent with the total kinetic energy loss rate $\approx 10^{27}$ erg s$^{-1}$ estimated for the whole  Sun assuming an asymptotic wind speed of 500 km s$^{-1}$ \citep[see, e.g., ][and references therein]{raouafi2023}, once the difference between this nominal downstream speed and the measured flow speed $v \approx 122$ km s$^{-1}$ near the coronal base is taken into account. Together with the scaling relation (\ref{eq:scaling}), these estimates strongly support a reconnection-driven solar-wind scenario in which the \corr{nonsteady} coronal plasma outflow is composed predominantly of small-scale jets, each ejected by a localized impulsive magnetic reconnection event in the low atmosphere, which propagate through a much slower background coronal plasma without significant mass and momentum exchange between the two systems. The latter condition is crucial for the described scenario and requires definitive verification in future studies.

\section{Conclusions}
\label{sec:conclusions}

We have conducted an in-depth quantitative investigation of the bulk plasma outflow from a polar coronal hole as represented by a sequence of 1,801 high-resolution, high-cadence EUV images above the north solar limb. To our knowledge, this is the first statistical survey reporting detailed probabilistic properties of the flows and evaluating their physical and geometric parameters based on a large set of automatically detected flow events. Below is a summary of our main conclusions.

\begin{enumerate}
  \item The upward plasma transport above a polar coronal hole is highly structured and contains numerous embedded flows.
  \item The average flow speed estimated using the surfing transform technique is 122 $\pm$ 34 \kms. The local speed varies significantly with position along the limb, likely reflecting the nonuniform photospheric magnetic structure at the base of the coronal hole. 
  \item The outflows consist of transient quasi-periodic bursts, with a repetition period of about 10 to 30 minutes and a characteristic burst duration of about 5 minutes. The typical transverse size of the individual flows is about 3--4 Mm, which is consistent with our earlier estimates \citep{kumar2022}.
  \item The spatial and temporal scales of the collimated flows obey broadband power-law probability distributions occupying a continuum of scales from the smallest jetlets to full-size jets, suggesting that the physical mechanism responsible for solar jets is inherently self-similar.
  \item The volume and kinetic-energy distributions of the flows are consistent with a scenario in which the structured coronal outflow is formed by a large number of impulsive reconnection events within the coronal hole, described by a wide range of spatial and temporal scales presumably imposed by the photospheric driver.
  \item Assuming that the multiscale jets and the inter-jet medium have comparable filling factors, the estimated occurrence rate of the analyzed transient events appears sufficient to produce the net mass and energy outflow for the nonsteady fast solar wind. 
    
\end{enumerate}

Overall, our smoking-gun findings add strong circumstantial support to the view that the \corr{nonsteady component of the} fast solar wind originates from small-scale \corr{impulsive} interchange reconnection in the low corona, as has been proposed earlier \citep{raouafi2023}. 
\corr{The quasi-steady component, on the other hand, may be driven by quasi-continuous interchange reconnection that adds heat and wave-associated momentum and kinetic energy to the plasma \citep{wang2022}.} 
More comprehensive high-resolution observations extending further into the corona, supported by high-resolution MHD simulations with full coronal thermodynamics, are needed to explore \corr{and quantify the contributions of impulsive and continuous reconnection, the global extent and persistence of the resulting outflows, and their ultimate connection to the heliosphere.}

\vspace{0.5cm}

SDO is a mission for NASA's Living With a Star (LWS) program. We thank J.\ Drake, S.\ Gibson, and A.\ Sterling for useful discussions. VMU has been partly supported through the Partnership for Heliophysics and Space Environment Research (NASA grant No.\ 80NSSC21M0180). PK is supported by an NSF grant (\#2229336) and NASA's Internal Scientist Funding Model (H-ISFM) program. CRD also received H-ISFM support. The Surfing Transform code was developed by VMU and is available upon request.


\end{document}